# Cloud Forensic: Issues, Challenges and Solution Models

Sayada Sonia Akter
Department of Computer Science
& Engineering
*United International University*
*United City, Madani Avenue,*
*Badda, Dhaka 1212, Bangladesh.*
sakter212039@mscse.uiu.ac.bd

Mohammad Shahriar Rahman
Department of Computer Science
& Engineering
*United International University*
*United City, Madani Avenue,*
*Badda, Dhaka 1212, Bangladesh.*
mshahriar@cse.uiu.ac.bd

ABSTRACT—Cloud computing is a web-based utility model that is becoming popular every day with the emergence of 4th Industrial Revolution, therefore, cybercrimes that affect web-based systems are also relevant to cloud computing. In order to conduct a forensic investigation into a cyber-attack, it is necessary to identify and locate the source of the attack as soon as possible. Although significant study has been done in this domain on obstacles and its solutions, research on approaches and strategies is still in its development stage. There are barriers at every stage of cloud forensics, therefore, before we can come up with a comprehensive way to deal with these problems, we must first comprehend the cloud technology and its forensics environment. Although there are articles that are linked to cloud forensics, there is not yet a paper that accumulated the contemporary concerns and solutions related to cloud forensic. Throughout this chapter, we have looked at the cloud environment, as well as the threats and attacks that it may be subjected to. We have also looked at the approaches that cloud forensics may take, as well as the various frameworks and the practical challenges and limitations they may face when dealing with cloud forensic investigations.

Acronyms—

| | |
|---|---|
| NFAT | Network Forensics Analysis Tool |
| CAGR | Compound Annual Growth Rate |
| NIST | National Institute of Standards and Technology |
| DDoS | Distributed Denial of Service |
| CSP | Cloud Service Provider |
| CSC | Cloud Service Customer |
| CSN | Cloud Service Partner |
| CSA | Cloud Security Alliance |
| APT | Advanced Persistent Threat |
| VMs | Virtual Machines |
| VTC | Virtualized Trusted Computing |
| SLA | Service Level Agreement |

# Table of Contents

# 1 INTRODUCTION

Cloud Computing is increasingly becoming one of the most advanced technologies, with a strong future for companies and organizations in the coming years. The era has experienced a revolution in cloud computing which not only caused many to consider the concept as a new Information Technology architecture, but also given it a reputation of being the most fast changing and industry-altering technologies since the invention of computers [1]. In 2020, cloud computing services produced more than $300 billion in sales. Considering the most recent cloud computing data, it's a fair bet that this business will only grow in the next decade [3]. Moreover, this massive growth has transformed how IT might be used to access, administer, develop, and offer services [2]. Indeed, using cloud computing in enterprises may lower the cost of IT [3]; which makes us feel that money is one of the key reasons why cloud computing is considered a quickly expanding technology.

On the other hand, the previously noted lightning-fast development in adoption of cloud computing has resulted in a scenario in which cloud settings are now seen as a foreign atmosphere for the implementation of cybercrime. This circumstance has also resulted in the emergence of completely new legal, organizational, and technological obstacles. At this moment, it is relevant to mention the considerable number of attacks that have an influence on cloud computing as well as the fact that cloud-based data processing is carried out in a manner that is decentralized; in fact, in addition to these factors, many people have voiced concerns related to how a comprehensive digital investigation can be undertaken in environments that are hosted by cloud providers [1]. Generally, it is vital to undertake investigations independently, without the need to depend on a third - party provider. But things are different in virtualized environment, where this process remains challenging. These complications stem from the fact that cloud services, who fully determine what takes place in the environment, keep control of the sources of evidence. Consequently, it is still hard to determine who is responsible for what. In addition, customers are still unable, at least to some degree, to proactively acquire data ahead to the occurrence of an event [4]. Thus, ensuring that forensic readiness has been achieved prior to conducting digital investigations, will lower the amount of money as well as time that would be required on the investigations. As per Market Research Media [14], the world - wide cloud computing industry has been forecasted to grow by a compound annual growth (CAGR) of 30% though the 2025; at this point, many feel that the market will be worth nearly $270 billion. This figure effectively demonstrates the expansion of the cloud computing sector as well as the rapid growing number of users in the cloud around the world. This expansion will directly result increase the number of cyberattacks. Although cloud forensic has a significant number of challenges, there are currently no regulations, processes, or standards established specifically for cloud forensics [5].

This chapter's objective is to provide the readers with an overview of Cloud Environment as well as the security challenges related with it. We also provide an analysis of a few prominent cloud-based cyberattacks. This sheds light on the need of cloud-based forensics as well as on the challenges associated with it. The National Institute of Standards and Technology (NIST) recently compiled a list of sixty-five cloud forensic challenges [6]. We go further to identify potential solutions to these challenges by analysing different cloud-based forensic models suggested by the authors. Finally, this research study may be regarded as an endeavor to encourage continuing research in this sector and build cloud forensic-capable systems.

The readers were given an introduction to cloud computing in the first portion of this chapter. The second section focuses more on cloud environment, including its characteristics and the standards that govern them. In the third section, we evaluated a few prominent cloud-based cyber-attacks and offered our observations on the reasons why these events occurred, which makes it clearly evident why it is essential to have cloud forensic in order to be forensically ready. The readers will have a fundamental understanding of cloud forensics after reading the fourth and sixth sections. In the seventh section, readers will get a grasp of Cloud Forensic Challenges. In the eighth section, we have evaluated numerous models suggested by the authors to address a number of significant challenges listed by NIST. In section nine, we offered an overview of the cloud forensic tools that are now accessible. Last but not least, we brought our review chapter to a close by including sections ten, eleven, and twelve that discussed potential opportunities and directions for future research.

## 2 CLOUD COMPUTING ENVIRONMENT

Cloud computing, as described by the National Institute of Standards and Technology [NIST], has "revolutionized the techniques through which digital data is stored, processed, and transported. It enables ubiquitous, accessible, on-demand network access to a shared pool of customizable computer resources (e.g., networks, servers, storage, applications, and services) that can be instantly supplied and released with minimum administrative effort or service provider contact" [6].

On-Demand Computing, or "Cloud Computing," is a kind of computing that relies solely on the Internet. Computers and other electronic devices may now be supplied with data, information, and other necessary shared resources on-demand using this computational technology. It is possible to store and process data in many third-party datacenters by providing cloud computing and storage options to numerous customers and businesses. As converged infrastructure and shared services are at the heart of this new technology, therefore, this technology separates information resources from the underlying infrastructure and delivery mechanisms.

### 2.1 Characteristics of Cloud Computing

Sharing resources is essential to cloud computing's goal of achieving coherence while also achieving cost efficiencies. Cloud computing allows businesses to execute their applications in a more efficient manner since resources are made accessible depending on the needs of the application. This technology was developed to produce systems that are efficient in terms of cost, allowing businesses to operate their processes without encountering any IT roadblocks. In cloud computing, a method known as virtualization was developed to partition a physical device into one or more virtual devices. This allowed efficient use of available resources. Computing in the cloud is comparable to computing on a grid. usefulness of this technology's may be associated with a number of factors, including:

o Self Service on Demand: the cloud service Consumers can use various services on demand without the help of the service providers.
o Pooling Resources: When resources are pooled, users may be served in one location or several locations, based on what is most beneficial for them individually.
o Wide Network Access: Resources related to Computing are sent across the network. Consumers can use the service once there is a network connection in their environment.
o Scalability: The resources can be set up without the help of service providers and can be rapidly scaled up or down to fit the requirements of the user.
o Measured service: The cloud architecture is able to make use of the appropriate strategies despite the fact that the computer resources are pooled and shared across a number of different participants.
o Maintenance: Because of its accessibility, maintenance of cloud apps is quite straightforward.
o Multi-tenancy: multi-tenancy is a useful feature that allows for cost and resource sharing.

### 2.2 Cloud Computing Standards

The international standard ISO/IEC 17788 defines a wide variety of service models for cloud computing. Within the community of cloud computing, three primary standards have been developed [24]- Service, Deployment, and Role. To begin, the cloud computing industry has relied heavily on three service models to classify cloud services, as shown in Figure 1.

o Infrastructure-as-a-Service (IaaS): IaaS is an acronym for infrastructure as a service. It is a kind of cloud computing that provides basic computing, network, and storage capabilities to customers on request, through the internet, on a basis of pay-as-you-go.
o Platform-as-a-Service (PaaS): When it comes to developing, running, and managing applications, PaaS is a model of cloud computing that provides customers with a complete cloud platform that includes hardware, software, and infrastructure without the cost, complexity, and inflexibility that are typically associated with building and maintaining that platform on-premises.

o   Software-as-a-Service (SaaS): Software-as-a-Service framework makes it possible to provide software applications to end customers in the form of a service. It is a term used to describe a piece of software that has been installed on a host service and is available to users through the internet.

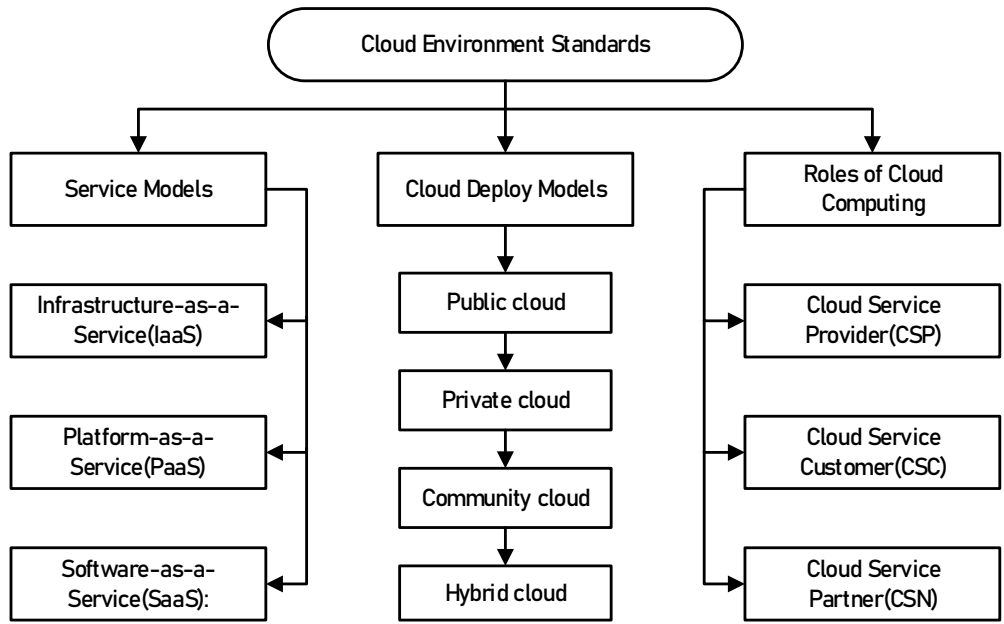

*Figure 1: Cloud Environment Standards.*

Cloud computing can be broken down into four categories: public, private, community, and hybrid:

o   Public cloud: Cloud services are accessible to every cloud service client, and the cloud service provider is in charge of the resources.
o   Private cloud: Cloud services are used by a specific cloud service client, who also retains full control over the resources made available via such cloud services.
o   Community cloud: A cloud environment that is shared among several entities and belong to the same community is referred to as a "community cloud." It is a cloud Infrastructure environment that provides the same level of privacy and security as a private Cloud.
o   Hybrid cloud: A hybrid cloud, also known as a cloud hybrid, is a type of cloud computing system that combines an on-premises datacenter, also known as a private cloud, with a public cloud. This makes it possible for users' data and applications to be shared between these two types of clouds.

At last, three distinct functions for cloud computing have been outlined- Cloud Service Provider (CSP), Cloud Service Customer (CSC), and Cloud Service Partner (CSN).

o   Cloud Service Provider (CSP): A third-party organization that offers platform, infrastructure, application, or storage services through the cloud is known as a cloud service provider (CSP).
o   Cloud Service Customer (CSC): A subscriber to a cloud service and a user of a cloud service are both examples of cloud service consumers. Within the limitations of a private cloud, all of the users and the service providers belong to the same company.
o   Cloud Service Partner (CSN): A partner that allows people to access the public Cloud services via their own interface. They are also responsible for billing and providing administrative assistance for billing-related matters.

*2.3   Threats in Cloud Environments*

Users of cloud computing do not know the precise location of their sensitive data since Cloud Service Providers (CSPs) operate their data centers in a variety of locations throughout the world; as a result, users are put in a vulnerable position regarding their safety. Standard security measures such as host-based antivirus software, intrusion detection systems and firewalls and do not offer enough protection for virtualized systems because of the rapid spread of threats that occur in virtualized settings. Per the Walker [25], the Cloud Security Alliance (CSA) has compiled and released a list of the top 12 threats that have been found to be associated with cloud computing. These potential issues are stated in Table 1. Out of all these potential dangers, the compromising of data has been pointed out as the most significant security problem that has to be addressed.

*Table 1: CSA'S Top 12 threats.*

| Threat no. | Threat name |
|---|---|
| 1 | Data breaches |
| 2 | Compromised credentials and broken authentication |
| 3 | Hacked interface and Application Program Interfaces |
| 4 | Exploited system vulnerabilities |
| 5 | Account hijacking |
| 6 | Malicious insiders |
| 7 | The Advanced Persistent Threat (APT) parasite |
| 8 | Permanent data loss |
| 9 | Inadequate diligence |
| 10 | Cloud service abuses |
| 11 | Denial-of-Service (DoS) attacks |
| 12 | Shared technology, shared dangers |

Several authors [26]-[32] came to the conclusion that the difficulties highlighted in Table 2 are preventing solutions from being established for threats. They consider these challenges to be the gaps for threat remediation and feel that they must be addressed in further study. It's interesting to note that a lot of these problems are trust concerns, which may result in cyber-attacks. These difficulties may be addressed from both directions [33]- Cloud service providers do not trust their consumers and believe that their customers may bypass existing privacy rules in order to carry out phishing and malware attacks while using the cloud services that the cloud providers offer. On the other hand, clients who use the cloud have a variety of additional trust difficulties. For instance, service providers may constantly attempt to disguise their own corporate rules for recruitment, and the incapacity of service providers to monitor their employees may result in an attack by malicious insiders. Images obtained from unreliable sources might leave a back door open for attackers to use. It is possible for cloud providers to become self-interested and greedy, and as a result, they may employ storage space with a lesser security level than was agreed upon, which may result in Cross VM Side channel attacks.

*Table 2: Defects in the process of threat remediation.*

| Security Threats | Challenges while implementing threat remediation |
|---|---|
| Misuse and Malicious Application of Cloud Computing | o Because of privacy restrictions, cloud service providers are unable to provide real-time monitoring.<br>o The interests of many stakeholders do not necessarily align in the same manner. |
| Insecure Application Programming Interfaces | o Inability to audit events connected to the usage of APIs.<br>o Incomplete data from the logs to allow the reconstruction of company's strategies. |
| Malicious Insiders | o Service providers may sometimes attempt to conceal their own corporate policies while hiring company employees.<br>o The solutions are implemented after the occurrence has already taken place, which is too late.<br>o The incapacity of cloud service providers to monitor their staff members. |

| Vulnerabilities Existing in Shared Technology | o Elements that are shared have never been intended for a high level of fragmentation.<br>o Competitors in the business world use distinct virtual machines hosted on the same physical machine.<br>o The presence of both- a retail and a manufacturing sector together. |
|---|---|
| Loss/Leakage of Confidential Data | o Lack of trust in the cloud providers since they could act in their own self-interest and keep data in a less secure location than was agreed upon.<br>o Procedures that have not been thoroughly evaluated, weak policies, and practices for the preservation of data.<br>o A poor understanding. |
| Hijacking of Accounts, Services, and Traffic | o Faster evolution of cloud computing results in the creation of new security vulnerabilities.<br>o Current method of managing digital identities is insufficient for use with hybrid cloud environments. |
| Unknown Risk Factors | o The reluctance of cloud service providers to give log and audit data; and procedures regarding cloud security.<br>o Lack of honesty and transparency. |

On the other hand, based on the findings of research conducted by relevant academics and the annual report of the Cloud Security Alliance (CSA) [7], there are significant dangers to the privacy security risk are shown in Figure 2.

o Identity authentication and Access control: Cloud computing requires an enormous number of resources, which significantly increases the administrative difficulty of managing access controls and identity verification.
o Privacy data security: As a result of the method of service outsourcing, the security risk associated with cloud privacy is very prominent. This risk includes challenges with data disclosure, privacy disclosure, access rights management, and data deletion.
o Multi-tenant and cross-domain sharing: multi-tenant and cross-domain sharing requires the assurance of multi-tenant isolation and multi-user security. When a domain is crossed, service authorization and access control become more difficult to manage, and the process of trust transfer among two cloud computing organizations needs to be reevaluated.
o Virtualization security: Although service providers have designed and implemented isolation techniques for virtual machines, it is impossible to completely avoid attacks between virtual machines. Migration of machines that are virtualized, will also produce changes in the security domain. Despite these efforts, attacks between virtual machines can still occur.
o Vulnerabilities of System Security: Because of the complexity of a cloud computing system, many service providers provide various management and service levels. As a result, the risk associated with using the cloud will grow if there are any security flaws.
o Advanced Persistent Threat (APT): it refers to an infiltration and attack that have been prepared on a cloud computing system. This system has developed certain underground interest chains.
o Wrong application of cloud service: Misusing cloud computing may result in difficulties for customers, service providers, or third parties, and illegally using cloud services will result in severe consequences.
o Unviability of Cloud Service: Denial-of-service attacks have become a significant security target for cloud service providers as a result of the fact that many security incidents are expressed as the inability of cloud computing services to be accessed.
o Threats generated by Insiders: It is common for the security policy to be ineffective due to the accidental or purposeful disclosure of sensitive information by service provider insiders. This has developed into a major concern for cloud computing security.

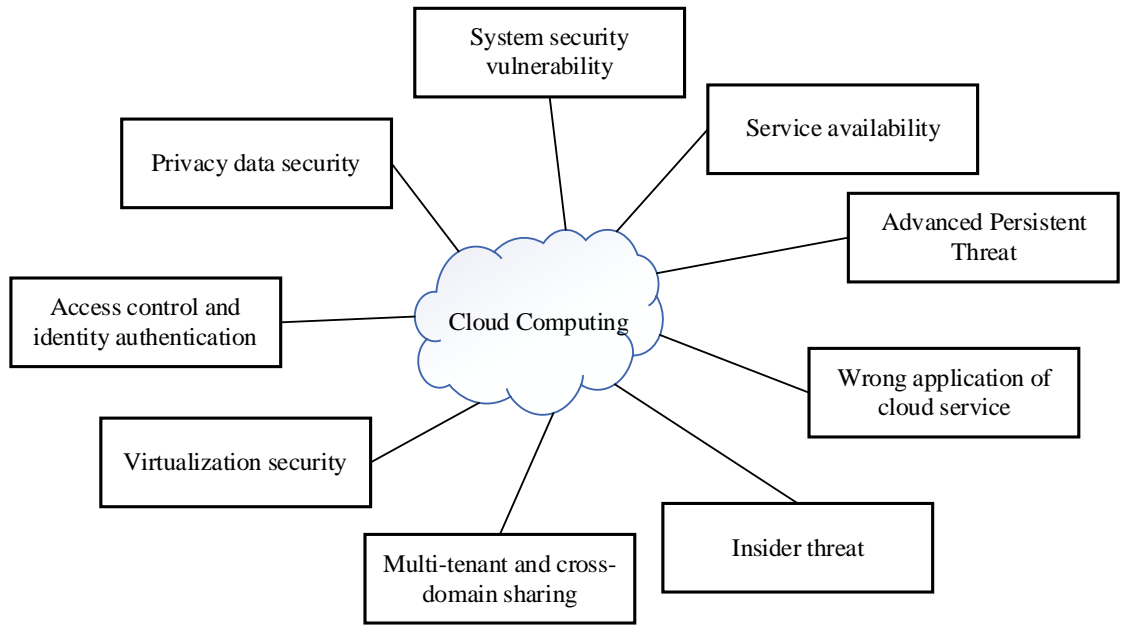

*Figure 2: Privacy security risk in cloud computing.*

## 3 TAXONOMY OF CLOUD BASED CYBER-ATTACKS

Any cloud-based service can be vulnerable to security breaches due to a number of factors, including misconfiguration, unauthorized access, insecure interfaces or application programming interfaces (APIs), account hijacking, or even malicious insiders. In this part of the chapter, we will discuss a few well-known cloud-based cyberattacks that caused businesses to reevaluate their security measures.

*Accenture Multiple Data Breaches*: In 2017, Accenture let at least four of its AWS S3 storage buckets to remain open, which resulted to a privacy compromise that exposed sensitive information such as unrestricted authentication details, API data, digital certificates, decryption keys, user data, and meta information [34].

In August of 2021, Accenture was hit by another attack, carried out by the LockBit ransomware. During audits of the company's fourth and final quarter in 2021, the organization had the expertise this time to identify the intrusion.

Because of the data breach that occurred in 2021, Accenture anticipated that chain attacks were being carried out on client systems. These attacks involved misconstrued critical systems, accidental exposure, and further malware infections. The perpetrators of the assault, the LockBit ransomware itself, claimed that they were able to steal 6 terabytes worth of data as a result of the attack and were demanding a ransom payment of $50 million.

*Verizon Attacks*: A misconfigured AWS S3 on the part of Nice Systems, a third-party partner of Verizon, led to the accidental disclosure of customer personally identifiable information in 2017 [35]. Because of an oversight made by Nice, the attack was able to take place. This fault allowed more customer call data to be captured. In the year 2020, Verizon discovered 29,207 security events, of which 5,200 were verified to be breaches. The telecommunications behemoth fell victim to distributed denial of service assaults (DDoS). Each attack was powered by social engineering and client-side web app infections, which led to server-side system vulnerabilities.

The pandemic-induced remote productivity model is the key cause for the development of cyberattacks as well as the emergence of weaknesses in the system. The group classified these attacks as being the consequence of mistakes made by the 'human factor,' which is a byproduct of social engineering.

*Kaseya Ransomware Attack*: A significant cyberattack was launched against the unified remote monitoring and network perimeter security technology used by IT solutions company Kaseya in the month of July 2021 [36]. A ransomware operation on the supply chain had the objective of stealing administrative control of Kaseya services from managed service providers and the consumers they serve downstream. The incident rendered the company's

Page **8** of 23

SaaS servers inoperable and disrupted on-premise VSA systems that Kaseya clients in ten different countries were using. The active response that Kaseya took to the attack consisted of quickly notifying its clients. The firm released the Kaseya VSA detection tool, which enables corporate customers to examine the VSA services they utilize and monitor endpoints while looking for indicators of threats.

*Cognyte*: Cognyte, a large cybersecurity research company, made a mistake in May 2021 that resulted in their database being left unprotected and without authentication, paving the way for cyberattackers, that exposed 5 billion user data [37]. User credentials such as identities, email addresses, and passcodes, together with vulnerability data points inside their system, were compromised and disclosed. The data was accessible to the general public and was even included in web search indexes. Additionally, Cognyte's intelligence data, which includes information about other data breaches of a similar kind, was made openly accessible to users. The data was secured by Cognyte over a period of four days' work.

*Raychat*: In 2021, A vulnerability in the database configuration allowed almost 267 million usernames, passwords, email addresses, metadata, and encrypted conversations to become public [39]. The entire company's data was deleted as a result of a coordinated bot attack. The business was keeping its customer information on a MongoDB database that was improperly configured. A NoSQL is a kind of storage application that is widely utilized by application firms to manage large amount of user information. However, if the NoSQL system is configured incorrectly, it can expose thousands of files to risk. In this particular instance, the malicious party managed to practically stroll right into the main doorway of Raychat and then execute a bot operation, which completely wrecked the database. The flaw was discovered by specialists while using open-source search tools that are utilized in the process of searching for gadgets that are linked to the internet. A large number of NoSQL databases, such as Mongo, are targets for BOT cyberattacks performed by malicious parties that scan the internet for open and unprotected DBS [databases] and delete their data, with just a ransom demand left behind.

*Civicom*: Amazon Simple Storage Service (Amazon S3) is a cloud-based storage solution that is scalable, has high transfer speeds, and is web-based. The service is created for use on Amazon Web Services with the aim of performing online backup and archiving of data and applications (AWS) [73]. The S3 bucket was left available to the public without a password or any other form of security authentication in October 2021 by Civicom, which meant that the data could be accessed by anyone who knew how to find corrupted databases. According to the findings of experts, Civicom disclosed 8 terabytes of documents, which included over 100,000 different files. This was caused by one of Civicom's Amazon S3 buckets having an inadequate configuration. In January 2022, after a period of three months, Civicom managed to secure the bucket.

*FishPig*: Malware was injected into Magento stores using a distribution network attack that is directed against the FishPig distribution server [74]. FishPig is a Magento extension supplier that has over 200,000 customers and specializes in Magento optimizations and interfaces between Magento and WordPress. This intrusion resulted in a threat actor introducing malicious PHP code into the Helper/License.php file. The injected code configures Rekoobe, which is another malicious program that hides itself as a background task inside systems that have been compromised. The malware introduced to License.php would download a Linux binary from license.fishpig.co.uk whenever the Fishpig control panel was accessed within the Magento system. The downloaded file, with the filename "lic.bin," pretends to be a licensing asset but is actually the Rekoobe remote access trojan. after being executed, the trojan deletes all harmful files from the infected machine but continues to operate in memory, imitating a system service, as it waits for commands from its command and control (C&C) server. The "lic.bin" file that was downloaded was actually a Rekoobe remote access trojan disguised as a licensing asset. The trojan removed all malicious files from the compromised computer but remained active in memory, masquerading as a system service while it awaits instructions from its command and control (C&C) server.

Microsoft: Microsoft reported on January 22, 2020 that one of its cloud databases had been accessed in December 2019, exposing 250 million emails, IP addresses, and support case information. The vulnerable customer service database was made up of a cluster of five Elasticsearch servers. This is a technology that is intended to facilitate search operations. Each of the five servers appeared to be a mirror image of each other since they all contained the same data. Microsoft claimed a faulty network server was the root cause for the data breach. Because of the prominent nature of the target, this incident was among one of the most alarming cyber-attacks.

Following are a few observations that emerge from our investigation on root factors that contribute to the occurrence of most of the cyber-attacks on cloud computing systems:

- Misconfiguration: CSPs provide several levels of service to their customers, and these tiers are determined by the amount of control an organization requires over its cloud deployment. In order to achieve a higher level of cybersecurity, businesses need to design these installations according to the specifications of their operations. Unfortunately, the majority of firms do not have suitable cloud defense capabilities to guarantee the safety of these services, which leads to vulnerabilities in the deployment process. According to IBM [9], faulty server configurations are the root cause of 86% of all compromised data. Having knowledge of the particular deployment that the firm is using will make it easier to configure it according to the company's unique security requirements using the security tools that are supplied by CSPs.
- Compromised user Accounts: The most common reason for user accounts to be compromised is inadequate password policies. The majority of users that use cloud services do not have adequate password security because they either use weak passwords, reuse previous passwords, or do not frequently update their passwords. Therefore, users should get strongly encouraged to change their passwords on a regular basis, at least once every sixty to ninety days.
- API Vulnerability: Users are able to communicate and collaborate with their cloud-based computing platforms due to the availability of application programming interfaces (APIs) by cloud service providers (CSPs). These application programming interfaces come with a comprehensive documentation package, which enables users to better comprehend and use the APIs. However, cybercriminals are also able to get these documentations, and it may be used to attack APIs in order to gain access to sensitive data that is kept in the cloud and then to remove that data. Additionally, any flaws in the implementation and setup of these APIs will make backdoors available for cybercriminals to get access to sensitive information. By strictly adhering to the documentation, one can mitigate the risks of making any security mistakes during the implementation and setup of APIs. Additionally, organizations are required to carefully monitor the APIs' functionality in order to locate any potential vulnerabilities.
- Insider Threats: A malicious user may bypass an organization's security policies and cause sensitive data to be compromised even if the organization has implemented the most advanced cybersecurity environment possible. Because they may already have access to sensitive data, malicious insiders' actions are sometimes difficult to identify because of the access they have to such data. In point of fact, the number of security breaches that have occurred as a consequence of insider threats has experienced a substantial increase over the course of the last few years. Implementing strong access controls allows organizations to limit the amount of information that can be accessed by persons working inside their own organization, which helps mitigate the risk posed by insider threats.
- In today's cloud-driven interactions, users provide increasing amounts of their sensitive information to corporations, despite the fact that many of these corporations fail to appropriately safeguard these data. Because of the malicious techniques and technologies that are utilized to obtain private information, there are trust and confidentiality concerns. When a situation like this arises, authorities such as law enforcement, detectives, and system administrators turn to digital forensics for assistance in rearranging the sequence of events and locating traces of evidence that have been left behind.

## 4 Cloud forensic as a form of Digital Forensics

Cloud forensics is a process that uses digital forensics techniques to investigate incidents that take place in the cloud. This technique is used to identify those responsible for the incidents. The use of digital forensic best practices inside a cloud context is what we mean when we talk about cloud forensics. According to NIST, Cloud computing forensic science uses scientific concepts, technology practices, and established procedures to recreate prior cloud computing events by identifying, acquiring, preserving, examining, interpreting, and reporting digital evidence [6].

After a crime has been committed, the forensic process may then begin. It discusses a variety of investigation methods [40], techniques, and tactics for certain crimes, in addition to the gathering of evidence. The current method of acquiring evidence has two major flaws: a) the investigators need access to the suspect's stolen machine, and b) traditional data storage, information exchange, and communication channels have been expanded by the newest emerging web services. Both of these flaws make the method less effective. It is possible for law enforcement agencies to request information from service providers; however, they are often not permitted to do so from other countries. As a consequence of this, the field of digital forensics is very new and is just in the beginning phases of its growth [41].

Since 1999, a variety of strategies and frameworks have been developed, each consisting of a number of steps and phases, in order to successfully carry out digital forensic investigations. McKemmish [42] was one of the first researchers to describe forensic computing (actually creating the terminology digital forensics) as "the process of identifying, preserving, analyzing, and presenting digital evidence in a manner that is legally admissible." McKemmish was also one of the first researchers to coin the term "digital forensics." The four most important aspects (stages) of the forensic computing process are the identification, preservation, analysis, and presentation of digital evidence. During the identification phase of the investigation, the investigators are tasked with locating any and all potential sources of evidence. During the stage when preservation is taking place, the chain of custody needs to be kept intact at all times. The stage that needs expert testimony in a legal setting is the presenting stage, which contrasts with the stage that requires analysis, which comprises the extraction, processing, and interpretation of digital data.

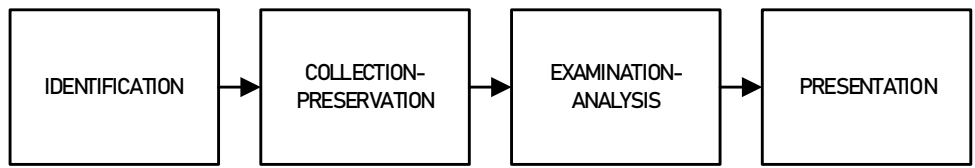

*Figure 3: Four Steps of Digital Forensics, proposed by authors*

## 5 DIMENTIONS OF CLOUD FORENSICS

NIST has put together a list of sixty-five cloud forensics difficulties. As a consequence of this, many industry professionals consider cloud computing to be an emerging platform for cyber-criminal activity [6]. The difficulties of how to perform a comprehensive digital investigation in cloud settings and how to plan to collect data ahead of time prior to the occurrence of an event are among these worries. Preparation of this kind would save money, effort, and time; thus, it is one of the primary concerns. On the basis of the evaluation of the relevant literature, the following framework is suggested: As shown in Figure 4, the structure comprises of three distinct dimensions [71].

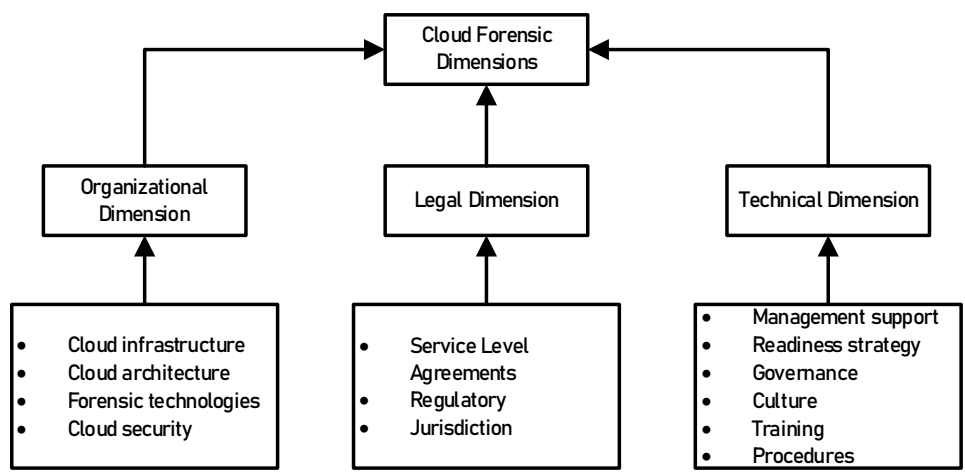

*Figure 4: Dimension of Cloud Forensics*

Page **11** of **23**

*5.1    Technical Dimension*:

Data gathering, live forensics, preventative measures, and virtualized settings are only some of the important tasks that have been completed. These tools are diverse from one another with regard to the deployment and service models that they use. Technical factors are- Support provided by Management, Forensic Readiness Techniques, Governance, Culture, Training, Procedures

*5.2    Legal Dimension*

This is necessary to assure user privacy and that forensics procedures do not violate any laws or regulations in data center jurisdictions. Legal factors are- Service Level Agreements, Regulatory, Jurisdiction

*5.3    Organizational Dimension*

Organizational Dimension: Both the cloud service provider and the cloud service consumers are members of the same organization. When a cloud service provider makes its services available to users, the scope of the investigation expands. This study involves the participation of three distinct outside parties, these are- Cloud infrastructure, Cloud architecture, Forensic technologies, Cloud security.

## 6    CLOUD FORENSIC PROCESS FLOW

The traditional steps of digital forensics, which can also be used in cloud forensics because of the different service and deployment models [72], are:

*6.1    Identification*

Reporting on malicious activities is a form of identification. This occur when a person or a CSP authority files complaint on some incident. This process has two different types of identifications: identifying what happened and identifying what was found.

*6.2    Evidence collection*

According to forensic standards [44], the investigator gathers bits of evidence from cloud service models like SaaS, IaaS, and PaaS without putting the evidence's integrity at risk. SaaS service model looks at each user's data through log files like error log, access log, data volumes, application log, transaction log etc.  The IaaS service model looks at raw machine files, system logs, backups, storage logs, and other types of data. The PaaS service model looks at the data of application-specific logs though the API, operating system exceptions, malware software warnings, etc. All the evidence that has been gathered must be kept safe so that it can be used in future investigations. This can be done by sending court order to cloud service providers. It's possible that keeping data safe will need a lot of storage space. The investigator talks about the rules for keeping data private and safe [45].

*6.3    Examination and analysis*

The analyst reviews the evidentiary information gathered by certain forensic tools in the previous stage. To get a logical conclusion, the criminal data is integrated, correlated, and assimilated. The analyst examines data from physical and logical files, regardless of where they are stored. It may be necessary to share the testimony with law police, a victim agency, or a person.

*6.4    Preservation*

All obtained evidence must be carefully stored and not tampered with for further investigation. Because the information is spread over many geographical locations, all log files must be saved.

*6.5    Presentation and reporting*

In the end, the investigator will put up a formal, well-organized report regarding the findings of the case so that it may be presented in a legal setting. The course of the cloud forensic method is shown in Figure 5:

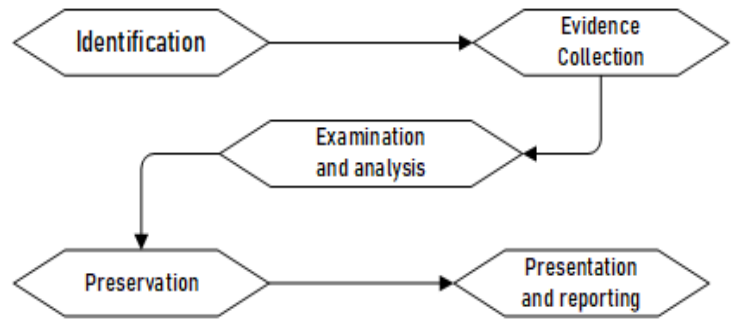

*Figure 5: Cloud Forensic Process*

## 7 CLOUD FORENSIC CHALLENGES

The cloud forensics faces a varieties of challenges. We have discussed a number of difficulties associated with cloud computing forensics [Figure 6] here:

o   Logs format unification: Since the cloud computing environment consists of a huge number of servers, each of which has its own log format, this will make the investigation process more difficult. Because of the large number of servers that make up a cloud system and the many geographic locations at which these servers are located, each location operates under its own unique time zone.

o   Intelligence processes for real-time investigation are often not possible in the cloud environment.

o   Data integrity and evidence preservation.

o   Lack of terms and conditions in SLA: Investigations have shown that the service level agreement (SLA), which should include the terms and conditions that govern the relationship between the user and the cloud service provider, is lacking terms and conditions. These bullet points need to include significant language associated with the investigation of cloud forensics [47].

o   Lack of forensics skills, particularly at the cloud computing level [48].

o   There is a lack of co-ordination in cross-national data access and sharing, lack of international coordination and regulatory mechanisms. Especially when cloud forensics relies on gathering evidence from servers in several locations.

o   To preserve evidence safely, strong encryption technologies are necessary [49].

o   CSP reliance: Investigators rely on CSP to get logs. Other than CSPs, there is no proof linking a certain information file to a specific suspect.

o   CSPs have agreements with other CSPs to utilize their services, which might result in data integrity and confidentiality being compromised in some situations.

o   Forensics in the cloud are dependent on client-server communication in order to maintain their integrity and consistency. When data is transferred from the investigator's workstation to the cloud storage device, it is done via a public network. Maintaining the integrity of the evidence is an essential part of the investigative process.

o   It's possible that the evidence in Cloud Forensics will be destroyed. It becomes difficult to retrieve erased material and recreate it for use as evidence.

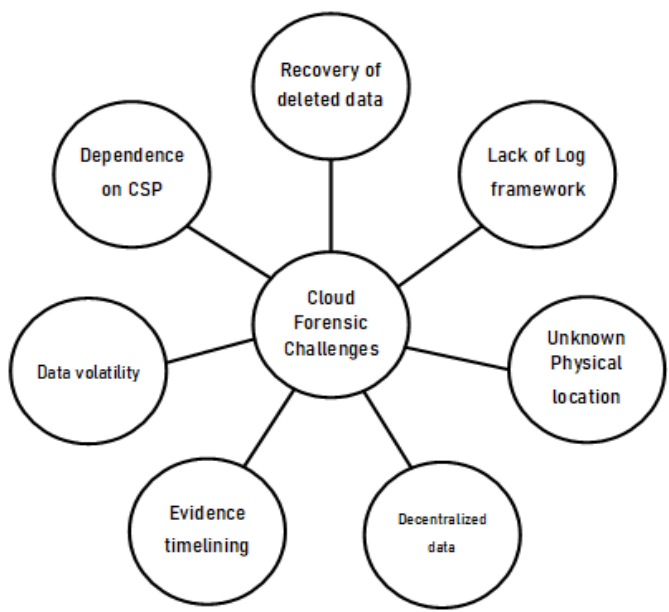

*Figure 6: Cloud Forensic Challenges*

"An on-demand network access approach that allows users to quickly provide and release shared computing resources (e.g., networks, server storage, applications and services) with no administrative effort or service provider contact is what NIST calls "cloud computing". Criminals, on the other hand, may take use of the properties of cloud computing in order to carry out illegal activities. Cybercriminals may exploit cloud computing as a topic, an object, or a tool [72]. Digital forensics in the cloud has a number of issues because of its diverse, virtualized, and distributed design [53]. In this way, new methods and tools may now be developed by researchers. The Forensics steps in a cloud environment are also shown in Table 3.

*Table 3: Challenges in Forensics phases in the cloud environment*

| Steps of Digital Forensics | Challenges in Cloud Forensics |
|---|---|
| Identification | Decentralized data, unknown physical location, jurisdiction, data duplication, encryption, dependency chain and dependency on CSP are all symptoms of this vulnerability. |
| Preservation | Evidence segregation, chain of custody, data volatility, distributed storage and data integrity. |
| Collection | Absence of specialized commercial tools, dependency on a CSP, lack of control over data erasure policy, many tenants, and several jurisdictions all contribute to a lack of usability. |
| Examination and Analysis | Log framework, evidence time lines, encrypted data, and Integration of evidence data. |
| Presentation and Reporting | Cloud model complexity and illiteracy, chain of custody, jurisdiction, compliance. |

## 8 CLOUD FORENSIC MODELS ADDRESSING THE CHALLENGES

In this part of the chapter, we classify various cloud forensic research models according to the challenges outlined by NIST. It has been possible to build and use several cloud forensics methodologies that are customized to certain types of deployment and service needs. Customers in PaaS and SaaS service models, for example, have no control over the hardware and must depend on the CSP for logs, but customers in IaaS have the ability to generate an image of the instance and retrieve the logs themselves. For deployment techniques, public cloud

Page **14** of 23

customers lack accessibility and privacy compared to private cloud users. Virtualization brings the private cloud idea closer to the old local access networks that were used in the past. On-premises private cloud forensics are essentially identical to conventional forensics when it comes to conducting forensic investigations (internally). A forensic analysis of a private cloud hosted off-site (external) is reliant on the CSPs and signed contracts. The most current research in cloud and digital forensics has been thoroughly examined, and the results are presented in this part in an in-depth study. This section includes ideas and theories from a variety of experts in digital and cloud forensics. To be clear, most of the works found are mainly focused with the investigation and resolution of cybercrime.

Case studies of widely used cloud storage systems such as DropBox, Amazon Cloud Drive, Amazon S3, and Google Docs were provided by a variety of authors fto address the challenges of storage forensics. They discovered the possible evidence artefacts in both the client-side as well as the server-side systems (cloud-native). Several articles [33, 70, 80, 81] made use of the application programming interfaces (APIs) provided by the service provider in order to gather forensic artefacts with the intention of overcoming the limitations of client-side analysis.

Authors [28-30] raised doubts about the conventional method of client-side storage analysis and switched the emphasis to cloud-side analysis instead. They brought up crucial concerns such as artefact modifications, partial replication, and cloud-native artefacts in their presentations. They also placed a strong emphasis on an acquisition that was based on an API in order to solve the issues with client-side artefact analysis. On the other hand, author [29] created the kumodd tool as a proof-of-concept. This had three logical layers: the user interface, the dispatcher, and the drivers. In addition, the authors validated their findings by using the cloud storage services Google Drive,9 Microsoft OneDrive,11 Dropbox,10 and Box23. Later, a Authors [30] pointed out a flaw in their previous work [29], that their kumodd tool lacked the ability to acquire cloud artefacts in their initial form due to the absence of API support. This flaw was emphasised in [30]. They developed a tool called kumodocs that is based on the Python programming language in order to investigate Google Docs artefacts. This tool is activated by a browser plugin called DraftBack24, which can replay the whole history of documents. The incompatibility of standard tools with StaaS (Storage as a service) applications was illustrated by Roussev et al. [28], who also presented 3 new forensic tools: kumodd, kumofs and kumodocs.

Geographic locations can be a difficulty during the investigation procedure if it takes place in a cloud environment. Authors [46] have suggested using a unified timekeeping system across all cloud entities as one of the potential solutions to the issue of time zones. This is one of the potential solutions that has been suggested. This has the benefit of establishing a logical order of events in terms of time.

Maintaining the integrity of the evidence is an essential part of the investigative process. There are many different solutions that have been proposed in order to improve the level of integrity of the cloud forensics investigation process. One such solution is [50], which suggests creating a digital signature for all of the collected evidences during the evidence gathering stage of the investigation, and then checking this signature on the other side prior to beginning the examination stage of the investigation. Hegarty [51] presented an additional solution to this problem, which offers a special framework for digital signature verification that enables forensics investigation of storage systems. The method was developed as an answer to this difficulty.

Authors [52] proposed scheme based on SecLaas that gathers logs for permanent storage and limits the possibility of manipulation by using Proof of Past Log (PPL). It safeguards the privacy of cloud users by storing the logs generated by virtual machines and giving forensic investigators access to the data they need to do their investigations. In addition to this, SeclaaS stores evidence of prior logs, which defends the logs' authenticity against dishonest investigators or cloud service providers. They claimed that the anonymity of the information might be maintained if the investigators accessed the logs via RESTful APIs. In addition to this, they suggested a modification of the Bloom filter as well as the Bloom tree for the integrity verification process in order to obtain increased performance in terms of both time and space.

The authors [57] used the irreversible nature of Blockchain to maintain the secrecy and authenticity of cloud logs, and they offered secure logging-as-a-service for the cloud environment. This was accomplished by using a distributed ledger. Steps such as extracting logs from a virtualized environment, creating encrypted log entries for each log using public key encryption, and storing encrypted log entries on the blockchain are all included in the system. Similarly, authors [58] suggested a public - key verification approach based on Blockchain that would use a third-party inspector to certify the authenticity of logs stored in cloud storage. The authors made use of homomorphic (calculating the hash of a composite block by adding the hashes of the separate blocks together) and one-way hash functions in order to produce labels for log entries, and they chose to store their data in a Merkel tree structure.

To address the Logs format unification challenge, a log format unification on cloud services with the use of the Cloud Auditing Data Federation (CADF) standard developed by the Distributed Management Task Force (DMTF) [54]. OpenStack, makes use of CADF event logging, and the authors have updated Apache CloudStack platform to make it forensically sound. They also analyzed the preexisting CloudStack platform in addition to the suggested CADF event model that was implemented.

Authors [55] developed cloud log assuring soundness and secrecy scheme (CLASS) for cloud forensic where public key of user was used to encrypt the logs. They used Rabin's fingerprint and bloom filter to avoid unauthorized modification of log. This approach reduced the time needed for verification.

An open-source framework was designed by the authors [56] using open-source tools such as- apache flume, apache kafka, ELK stack and apache spark for real-time and historical log analytics which can assist the system admins to monitor critical incidents while analyzing in real time.

Because of the virtualization, distribution, and dynamic nature of cloud systems, developing a cloud architecture that can enable forensics is a huge and difficult challenge. This challenge involves a large number of delicate legal, organizational, and technological challenges. Authors [59] offered an effective Cloud Forensics Investigation Model (CFIM) that can investigate crimes committed in the cloud in a manner that is both forensically sound and timely. The system is compatible with the idea of Forensic as a Service (FaaS), which offers a multitude of advantages to the process of doing digital forensics by using Forensic Server on the cloud side.

To ensure the trust and decreasing the dependency on CSPs, author [60] have offered a model in order to improve cloud forensics. In this model, both the forensic monitoring plane (FMP) and the forensic server have been developed. The forensics tool such as forensic toolkit (FTK) analyzer, E-Detection running at the top of the FMP will supervise both inbound and outbound links in the cloud system. The data that are monitored are encrypted bit by bit stream and stored on a separate forensics server that is positioned in crime site. The forensic tool monitors how cloud service models behave. Therefore, when an event takes place in a particular cloud, all of the actions that take place in that cloud, including network traffic, are forensically imaged. The image is encrypted again and stored in the forensic server which enables a reduction in the trust amount placed in the CSP. In the event that there was any kind of malicious behavior, the investigator may immediately connect to the forensic server using their user credentials, and they will be able to gather forensic evidence within a reasonable amount of time after the incident. In the meanwhile, if the investigator has any reason to suspect anything, they may make a data request to CSP and compare it to the information they have gotten from the forensic server.

A log assuring confidentiality mechanism suitable for use in Edge-Cloud Environment was suggested by the authors [61]. The investigators will attempt to retrieve the log files using this method. In the beginning, each and every log will be saved on one of the Edge nodes. After the segmentation has been saved to the centralized Cloud node and the distributed storage system, the users will log in. During the initial phase of the attack, the attacker will attempt to delete or steal data from the Edge node rather than the central store or the distributed storage system. In the second step of the attack, the attacker will attempt to steal data by attacking the Cloud node. Meanwhile, CSP will communicate with investigator so that they can investigate the data and store it. The investigator is able to retrieve the data from the central storage or distributed storage system, but the Edge node, which was either taken or destroyed by the attacker, is beyond his ability to recover the data. In addition, the data may be eventually retrieved from the distributed storage clusters with the assistance of the MIC network and index files.

Authors [62, 63] sought to solve the problems that arise while attempting to reconstruct events in a cloud context due to multi-tenancy and privacy violations, various heterogeneous data sources, and a huge number of events. They suggested frameworks and techniques for successful cloud event restoration based on the log aggregation algorithm Leader-Follower (LF) [62] and classical digital event restoration [63] respectively.

A Forensic Pattern-Built Technique was developed in 2018 by Juan-Carlos Bennett and Mamadou H. Diallo [66]. This approach is a semiformal architecture that is based on an object-oriented approach in the context of patterns. They have been using the NIST Forensics Framework throughout the whole process of collecting, reviewing, and analyzing the evidence. They developed the Cloud Evidence Collector in addition to the Cloud Evidence Analyzer so that they could collect more and better evidence of the network in the shortest period of time feasible.

Although it may not be commonly explored in earlier researches, log correlation is an important factor to take into account when doing analysis of cloud logs. For instance, Authors [43] concentrated only on the logging of certain programs and did not take into account the connection between the various logs. The prototype implementation that is described in Reference [64] was created for Windows event logs, and the safety of the centralized log server was not taken into consideration throughout its development. Kernel change [69] for forensically enabled cloud environment may be expensive because of the size of the cloud's hardware infrastructure, and it may not be practical for cloud service providers (CSPs) since it would necessitate the shutdown and restart of existing cloud services. The method that other authors [70] used was one in which the authors did not show any functioning prototype of their proposed notion and instead made the assumption that perhaps the CSP can be trusted completely.

Park [75] proposed a forensics framework that is a blockchain-based data storage and integrity management mechanism. Since blockchain maintains the integrity of data, this framework should be used to store forensic evidence. As a result of the fact that all of the blocks are linked to one another, the integrity of the data can be easily checked by using the hash value of the block that comes before it.

Protecting the logs is yet another crucial component of the process [76]. Cloud maintains a record of each, and every action taken inside its environment. On the basis of the log data, a forensics expert conducts the investigation into the incident. A forensics investigator's primary concern should be protecting the users' privacy wherever possible. This is necessitated by the fact that cloud computing is utilized continuously by plenty of users, all of whom share the same storage and processing network.

CSPs have to be available by the third-party investigators for inspection of logs. In order to solve this problem, there is a log block tag system [77] that is based on the Merkle Hash Tree.

Pourvahab [78] designed a blockchain system that operates in the cloud with IaaS concept. The forensic design of the blockchain, in which peers are scattered among different nodes, is used in this method for the purpose of gathering and storing evidence. It was intended that a secure ring verification-based authentication would be used for the purpose of safeguarding a device from being used by unauthorized parties.

Maintaining the integrity of the evidence is an essential part of the investigative process. There are many different solutions that have been proposed in order to improve the level of integrity of the cloud forensics investigation process. One such solution is [50], which suggests creating a digital signature for all of the collected evidences during the evidence gathering stage of the investigation, and then checking this signature on the other side prior to beginning the examination stage of the investigation. Hegarty [51] presented an additional solution to this problem, which offers a special framework for digital signature verification that enables forensics investigation of storage systems. This method was developed as an answer to this difficulty.

## 9 CLOUD SPECIFIC FORENSIC TOOLS

The development of digital forensic tools is still in the early stages due to the ever-changing nature of new technologies and the ever-increasing number of devices that utilize these technologies. The National Institute of Standards and Technology (NIST) produced a database of numerous digital forensics tools and their capabilities [67]. However, the category for cloud services only contains a set of six tools altogether. In this area, we classify forensics tools into two categories: cloud-specific solutions, and classic digital forensics tools. [68] There is a list of these instruments in Table 4.

*Table 4: Conventional Digital Forensic Tools.*

| Tools | Functions | Service Model/ OS |
|---|---|---|
| Digital Forensic Tools | | |
| DFF [25] | Instrument used in forensic investigation to detect, acquire, and store evidence while maintaining a chain of custody. | IaaS |
| EnCase Forensic [31] | A forensic solution in the form of a set of software that may be used to gather, preserve, analyse, and report on evidences in a manner that is acceptable to the court. | IaaS |
| Access Data Forensic Toolkit (FTK) [25] [31] | It is a collection of forensic tools, such as FTK imager for disc imaging, that may be used to analyse emails, carve data, and do other tasks. | IaaS |
| Wireshark [31] | Network protocol analyser | IaaS |
| Wild packets Omnipeek40 [21] | Enterprise-level software for the examination of network packets and protocols. | All |
| eNetwork Miner [10] | Open-source Network Forensics Analysis Tool (NFAT). | All |
| X-Ways Forensic [11] | Integrated forensic environment with a range of capabilities such as access to file system structures with deleted partitions, disc imaging and cloning, file and directory catalogue, and more. | SaaS |
| Cloud Specific Tools | | |
| FROST [38] | A forensic toolset built for the OpenStack cloud platform that can collect forensic evidence independently of a cloud service provider (CSP). | IaaS |
| PALADIN [12] | This digital forensic program offers more than one hundred helpful tools that may be used to analyse any potentially harmful content. | Windows and Linux |
| Kumodd [31] [59] | A forensic tool for cloud storage that can acquire cloud drives and take snapshots of cloud-native artefacts in formats such as PDF. | SaaS |
| e-Fencer [13] | It provides investigators with a straightforward interface for searching for data on whatever device they want. | Windows, Mac OS X, and Linux |
| Kumodocs [59] | Analysis software for Google Docs relying on Draft-Back, a browser plugin that replays the whole history of documents sitting in the Document folder. Draft-Back was developed by Google. | SaaS |
| Kumofs [59] | Tool used in forensic investigations for the gathering and examination of file metadata that is stored in the cloud. | SaaS |
| VNsnap [8] | Cloud-based snapshot tool designed for use with virtual network architectures. | IaaS |
| EnCase [15] | The ability to recover credentials from the hard disc is facilitated for the investigators by this. It enables investigators to conduct in-depth analyses of case files in order to collect evidence such as papers, images, and so on. | Windows |
| Crowdstrike [16] | <ul><li>Backing up virtual, physical and Cloud-based data centres.</li><li>Managing system errors.</li><li>Automatically detecting malware.</li></ul> | Windows and Mac |
| Xplico [17] | <ul><li>Data output into MySQL or SQLite database.</li><li>Reserves DNS search from DNS packages that contain input files.</li><li>Gives features like Port Protocol Identification (PPI) to assist digital forensics.</li></ul> | Linux |

|  | • Open source and supports IPv4 as well as IPv6. |  |
|---|---|---|
| SANS SIFT [18] | • Installable using SIFT CLI.<br>• Improves memory.<br>• Latest forensic methods and tools | Windows, Linux, Mac OS X, |
| Cloud Data Imager [33] | Innovative solution for the remote acquisition of cloud storage that primarily consists of two features: directory browsing and the creation of a logical duplicate of the folder tree that has been picked. | SaaS |
| LINEA [19] | A forensic instrument for the gathering of live network evidence from various web services. | SaaS |
|  |  |  |
| ForenVisor [68] | A piece of software in the form of a dynamic hypervisor that can do live forensic analysis. | IaaS |

**10 OPPORTUNITIES OF CLOUD FORENSICS**

Cost Efficiency: Comparing to Cloud Computing, which is more cost-effective and less expensive when deployed at a bigger size, Cloud Forensics is also less expensive when executed at a larger scale.

Abundance of Data: Due to the fact that the data is replicated and stored across multiple data centers and servers, it is possible that the data has not been completely deleted. This data may be available for forensics, and even if it has been deleted, it may still be possible to recover it. This is due to the fact that the data is replicated and stored across multiple data centers and servers.

Regulations and guidelines: Because cloud computing is still in its development era, it's the ideal moment for Cloud Forensics to establish the groundwork for their policies and standards.

Forensics as a Service: Forensics as a Service, also known as FaaS, is something that can be created, and once it is, it may be of great support in solving crimes linked to the cloud, as well as other types of cyber-crime investigations.

**11 FUTURE WORK**

Cloud forensics investigators must collect all evidence from a distributed cloud architecture. Future research can focus on building a reliable and effective framework for investigators to avoid errors in evidence collection and integration. Methods of machine learning, such as log correlation and meta-data analysis, are becoming an increasingly important focal point in cloud computing. Some stages of cloud forensics, such as federated learning, may benefit from the use of automation [65]. Cloud forensics datasets that include cloud meta-data attributes are not publicly available. The future includes a contribution to the cloud forensics dataset. Obtaining evidence data, protecting the naive user's privacy, and adhering to new legislation requires current legal knowledge. The researchers are moving in the direction of standardization as the ultimate objective of cloud forensics. The provision of an all-encompassing cloud forensics solution is one way to achieve standardization.

**12 CONCLUSION**

Nowadays Cloud computing is one of the most important issues in information technology. Everything is on the cloud and people can easily access it from anywhere through the internet. Therefore, all the giant organizations are using these services or are planning to move towards it. Any new development in the cloud will almost certainly uncover some previously unknown dangers. We believe this is mostly due to increasing competence in the field as well as growing acceptance of cloud computing within the information and communications technology sector. The field of cloud forensics is fraught with difficulties at every stage of the investigation process. This comprehensive literature review that is offered in the article demonstrates that cloud forensics is moving in the direction of evidence provenance. In addition to that, the study demonstrates how the various aspects of cloud forensics may have an effect on the forensics process. We have notified a number of novel cloud forensic taxonomy in this chapter that considers challenges, evidence acquisition logistics, analysis, trust, and legal implications. In addition to this, this chapter makes it abundantly clear that future research might concentrate on developing a dependable and efficient framework for investigators to use in order to prevent making mistakes in the process of evidence collecting and integration. An effective contribution to cloud forensics may be made through the intelligent gathering of prospective evidence. Despite the fact that trust problems cannot be eliminated entirely, the use of a system that is based on provenance may give the necessary level of confidence for the inquiry.